\documentclass[prb,twocolumn,showpacs,preprintnumbers,amsmath,amssymb]{revtex4} 
 
\usepackage{graphicx}
\usepackage{dcolumn}
 
\newcommand{\beq}{\begin{equation}} 
\newcommand{\eeq}{\end{equation}} 
\newcommand{\beqa}{\begin{eqnarray}} 
\newcommand{\eeqa}{\end{eqnarray}}

\def\up{\uparrow}
\def\down{\downarrow}

\def\bk{{\bf k}}

\def\de{{\partial}}

\begin{document} 
\title{Multiple Gaps and Superfluid Density from Interband Pairing in Iron Oxypnictides} 
 
\author{L. Benfatto$^{1,2}$, M. Capone$^{2,3}$, S. Caprara$^{2}$, C. Castellani$^{2}$, and
C. Di Castro$^{2}$}

\affiliation{$^{1}$ Centro Studi e Ricerche ``Enrico Fermi'', via Panisperna 89/A, 00184, Rome, Italy} 
\affiliation{$^{2}$ CRS SMC, CNR-INFM and Dipartimento di Fisica, Universit\a di Roma ``La Sapienza'', Piazzale A. Moro 2, 00185, Rome, Italy}
\affiliation{$^{3}$ ISC-CNR, Via dei Taurini 19, 00185, Rome, Italy} 

\begin{abstract}
We study a four-band model for the iron oxypnictides, in which the
superconducting properties are assumed to be determined by the interband
coupling between hole-like and electron-like Fermi sheets.  We show that
reasonable parameters can account for the angle-resolved photoemission
spectra showing multiple gaps in Ba$_{1-x}$K$_x$Fe$_2$As$_2$, and for the
temperature dependence of the superfluid density. At the same time, the
zero-temperature value of the superfluid density shows a conventional
scaling with the number of carriers.

\end{abstract}
 
\date{\today} 
\maketitle 

The discovery of superconductivity in the family of compounds characterized
by Fe-As layers\cite{kamihara} is intriguing and challenging. Moreover, 
at the moment they are the only materials with $T_c$ larger than
40K beside the cuprates. While some properties of these materials are
reminiscent of the cuprates (superconductivity induced by doping a magnetic
parent compound, layered basic structure), the differences are also
remarkable.

Among the peculiarities of the iron oxypnictides not found in cuprates we
mention the magnetic nature of Fe and the multi-band character of the
band structure. The effect of correlations seems to be weaker than for the
cuprates\cite{gabi}, even if also in the latter case the correlation
strength may be smaller than usually believed\cite{comanac}. At the same
time, the calculated electron-phonon coupling can not account for the high
critical temperatures\cite{boeri}, and some more exotic mechanism is
expected to be realized.

The initial efforts in the field focused on RO$_{1-x}$F$_{x}$FeAs and
RO$_{1-\delta}$FeAs (R being a rare earth atom), in which electrons are
doped in the parent ROFeAs. Even more recently, a new direction opened
thanks to the synthesis of Ba$_{1-x}$K$_{x}$Fe$_2$As$_2$ and
Sr$_{1-x}$K$_x$Fe$_2$As$_2$, doped by holes. The great advantage of this
class of materials is the availability of very large single crystals, which
allows for a much more complete and reliable experimental analysis.

Density functional theory (DFT) calculations of the band structure of
oxypnictides show that the Fermi surface (FS) is formed by disconnected
hole-like pockets centered around the $\Gamma$ point, and electron-like
pockets centered around the M points of the folded Brillouin zone of the
FeAs planes.  All the pockets essentially arise from the five d-orbitals of
Fe, three pockets belonging to the hole-like manifold and two forming the
electron-like one\cite{mazin}. Such a FS topology, along with the magnetic
character of the parent compound, motivated one of the first proposal, that
assigned to the interband coupling between the two manifold of FS's the
role of driving force behind the superconducting
phenomenon\cite{mazin,altri}.  From a microscopic point of view, the origin
of the interband pairing can be associated to spin fluctuations. Indeed,
the magnetic ordering vector of the parent compounds roughly connects the
two sets of hole and electron FS's, leading to an (approximate) nesting,
eventually giving rise to an effective pairing interaction.  On the other
hand, attractive phonon-induced intraband pairing (or the pairing between
FS's of the same nature) is likely to be relatively small.\cite{boeri}  In
this perspective, different nesting properties between the various FS
sheets can be expected to lead to different strengths of the pairing
interaction in the various bands.

Recent angle-resolved photoemission (ARPES) results for undoped\cite{feng}
and hole-doped $x=0.4$ BaFe$_2$As$_2$\cite{ding} clearly show that this can
indeed be the case. From a broad perspective, ARPES confirms the prediction
of DFT about the existence of distinct electron and hole-like
pockets. However, only two out of the three hole-like pockets may be
resolved around the $\Gamma$ point, with considerable different areas, and
a significant mass renormalization is found in all the bands with respect
to DFT.\cite{bande122} The superconducting (SC) gap opens on all the FS 
sheets at the same temperature, with a nodeless structure essentially
compatible with simple s-wave symmetry. More remarkably, the gap at the
smallest hole-like FS and the one at the electron-like sheets are almost
identical, while the largest hole-like FS has half of this gap. As we shall
see below, this effect is highly non trivial, and requires a proper
analysis of the multiband model and of the role of interband interactions.
The experimental situation is less clear for the electron-doped
materials. A nodeless SC gap at the $\Gamma$-FS has been reported for
NdFeAsO$_{0.9}$F$_{0.1}$\cite{liukondo}, but no information is available to
our knowledge about the gap in the electron-like pocket(s).

Somehow related to the issue of the multiband structure is the behavior of
the superfluid density $\rho_s$, both at zero and finite temperature.  The
$T=0$ BCS value of $\rho_s$ is indeed related to the filling and effective
masses of the various bands, while its temperature dependence probes via
quasiparticle excitations the presence of distinct SC gaps.  The aim of
this paper is to study a simple model which can capture the main effects of
the multiple-gaps opening in BaFe$_2$As$_2$ superconductors. We will show
that a reasonable description of ARPES results and superfluid density can
be achieved within a four-band model with realistic parameters, and that a
doping-dependent renormalization of the band structure should be
envisaged. Finally, we show that the calculated $T=0$ value of the
superfluid density is in reasonable agreement with the
experiments,\cite{uemura_ba,klauss_short,klauss_long,martin,li_optics} in
striking contrast to cuprates, where the anomalous scaling of $\rho_s$ with
doping can only be accounted for by including correlation
effects\cite{massimo_rhos,lara_pp}.

%
\begin{figure}[h]
\includegraphics[scale=0.3]{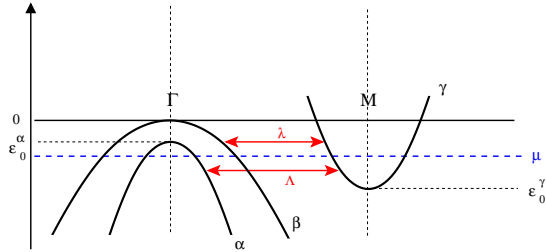}
\caption{(Color Online) Schematic of the multiband model used in this
  work. The two hole bands $\alpha,\beta$ are centered around the $\Gamma$
  point, the electron bands around the M point.}
\label{figband}
\end{figure}
%

We consider a four-band model with parabolic dispersion,
$\epsilon_i(\bk)=\epsilon_0^i\pm t_i (\bk-\bk_i)^2$, with the plus (minus)
sign for the electron (hole) bands and $t_i$ hopping parameter (we will
use, unless explicitly stated, units $\hbar=a=1$, $a$ being the lattice
spacing). For the two hole bands around the $\Gamma$ point, labeled as
$\alpha$ and $\beta$ (according to the notation of Ref.\
[\onlinecite{ding}]), $\bk_i=0$, while for the two electron bands $\gamma$
around the $M$ points $\bk_i=(\pi,\pi)$. As we mentioned above, the third
hole pocket predicted by DFT is not seen around the $\Gamma$ point,
probably because highly degenerate with the $\alpha$ one. Moreover, the two
hole pockets are slightly split in the undoped system.\cite{feng} Thus, we
choose the reference energy such that
$\epsilon_0^{\beta}=0$, while $\epsilon_0^\alpha$ and $\epsilon_0^\gamma$
are in general different from zero, see Fig.\ \ref{figband}. For simplicity,
we assume that the two electron bands are equal and isotropic around the M
points, even though both DFT and experiments show that they are slightly
elongated along the $k_x$ and $k_y$ directions. Within a BCS-like approach,
the pairing Hamiltonian can be written as:
\beqa
\label{bcs}
H=\sum_{i,\bk\sigma} \epsilon_i(\bk) c^+_{i,\bk\sigma}c_{i,\bk\sigma}+
\sum_{i\neq j \atop{\bk,\bk'}}V_{i,j}c^+_{i,\bk\up} c^+_{i,-\bk\down}
c_{j,-\bk'\down} c_{j,\bk'\up},
\eeqa
where $c_{i,\bk}^+$ creates an electron in the $i$-th band
($i=\alpha,\beta,\gamma$), and as we discussed above we neglect
intraband pairing. Moreover, the similar size of the FS for the
$\alpha$ and $\gamma$ bands suggests that the largest pairing acts between
these two FS sheets, which are strongly nested, while the $\beta$ and
$\gamma$ bands are less coupled.  These conditions are implemented in
(\ref{bcs}) by taking $V_{\alpha\beta}=0$, $V_{\beta\gamma}=\lambda$ and
$V_{\alpha\gamma}=\Lambda>\lambda$.  As it has been noticed in Ref.\
\cite{mazin,altri}, for inter-band mechanisms it is not necessary that the
interaction is attractive (at least at low energy), as it happens for
intra-band mechanisms. However, if $\Lambda,\lambda>0$ one sees that the
order parameters must have different sign on different kind of FS sheets,
i.e. $\mathrm {sign}\Delta_{\alpha,\beta}=-\mathrm
{sign}\Delta_\gamma$. The resulting set of self-consistent equations for the
absolute values of the gaps is a straightforward extension of the usual BCS
equations:
\beqa
\label{da}
\Delta_\alpha&=&\Lambda\Delta_\gamma 2\Pi_\gamma\\
\label{db}
\Delta_\beta&=&\lambda\Delta_\gamma 2\Pi_\gamma\\
\label{dg}
\Delta_\gamma&=&\Lambda\Delta_\alpha\Pi_\alpha+\lambda\Delta_\beta\Pi_\beta
\eeqa
where $\Delta_i$ is the gap in the $i$-th band, whose dispersion is now
$E_{i,\bk}=\sqrt{\xi_i^2(\bk)+\Delta_i^2}$, with $\xi_i\equiv
\epsilon_i-\mu$. The $\Pi_i$ are the particle-particle bubbles, given by
$\Pi_i=N_i \int_0^{\omega_0}d\xi ({1}/{E_i})\tanh\left({
E_i}/{2T}\right)$, where the pairing is only effective for states in a range
$|\xi|<\omega_0$ around the Fermi level and $N_i=1/4\pi t_i$ is the density
of states (DOS) of each band. Here we use $\omega_0=20$ meV, i.e. of the
order of the smallest Fermi energy in the different sheets.  Along with
Eqs.\ (\ref{da})-(\ref{dg}) we must solve self-consistently the equation
for the particle number $n=n_\alpha+n_{\beta}+2n_\gamma$, where $n=4$ is
the half filling.
At the present stage the agreement between the experimental dispersion and
the DFT bandstructure is hardly satisfactory\cite{bande122}, therefore we
extract the band parameters from the experimental data of Ref.\ \cite{feng}
for undoped BaFe$_2$As$_2$. Here one has $t_\alpha=90$ meV, $t_\beta=39$
meV and $t_\gamma=45$ meV, with a band splitting $\epsilon_0^\alpha=-10$
meV and $\epsilon_0^\gamma=-45$ meV. However, in a rigid-band picture, with
these parameters at the doping $x=0.4$ the electron pockets would be
empty. Moreover, as we shall see below in Fig.\ \ref{figgap}b, with these
parameters the gap values in the $\alpha$ and $\gamma$ bands would be quite
different, in contrast to what measured in Ref.\ \cite{ding}. We thus
proceed by assuming a band renormalization with respect to half-filling to
obtain $\Delta_\alpha\simeq \Delta_\gamma$, and
$\Delta_\alpha=2\Delta_\beta$ from Eqs.\ (\ref{da})-(\ref{dg}), where the
gap values depend only on the DOS and on the coupling.  The latter
condition is always realized for $\lambda=\Lambda/2$.  The former condition
instead can be obtained with slightly renormalized bands, i.e. we use
$t_\alpha=65$ meV, $t_\beta=25$ meV and $t_\gamma=60$ meV. Notice that if a
third hole pocket of the same size of the $\alpha$ one is present, even if
not explicitly observed by ARPES, this would just require to slightly
change the parameters to compensate the substitution of
$\Pi_\alpha\rightarrow 2\Pi_\alpha$ in Eq.\ (\ref{dg}).

%
\begin{figure}[htb]
\includegraphics[scale=0.3,angle=-90]{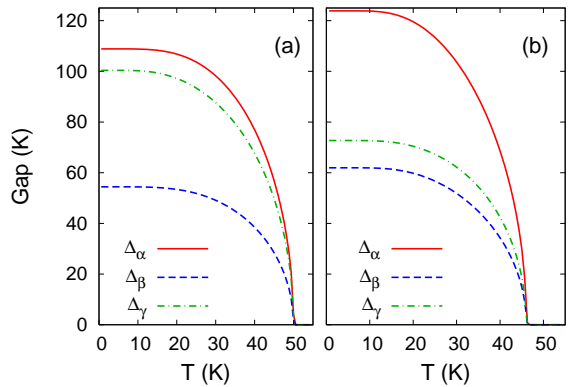}
\caption{(Color Online) (a): Temperature dependence of the SC
  gaps in the hole and electron bands. The parameter values are optimized
  to obtain the same gap value in the $\alpha$ and $\gamma$ bands (values
  in eV). Moreover, we used $\lambda=\Lambda/2$ to have
  $\Delta_\beta=\Delta_\alpha/2$. (b): SC gaps for the hopping parameters
  estimated from ARPES measurements of Ref.\ \cite{feng} at half-filling.}
\label{figgap}
\end{figure}
%

Once the DOS are fixed by the gap values, we tune the band
splitting according to the FS areas $A_i$ measured in Ref.\ \cite{ding}. In
particular, it is observed that
$A_\alpha/A_\beta=2/9$. Notice that if $\alpha$ and $\beta$ bands were
degenerate (or nearly degenerate) at the $\Gamma$ point this
$A_\alpha/A_\beta$ ratio would require a very large anisotropy between the
hole bands, $t_\beta=t_\alpha(2/9)$, which in turn would not allow the
condition $\Delta_\alpha\simeq \Delta_\gamma$ to be fulfilled.  Thus, to
justify the ratios of the FS areas we should assume a splitting of the hole
bands at the $\Gamma$ point, even larger than what measured 
at half-filling\cite{feng}. This enhancement of the splitting can be
associated to a different Hartree shift of the two bands, arising from a
particle-hole decoupling of the interband interaction. Imposing the FS
ratios suggested in Ref.\ \cite{ding} we determine
$\epsilon_0^\gamma=-71$ meV and $\epsilon_0^\alpha=-25$ meV (so that
$\mu\simeq-60$ meV).  In Fig.\ \ref{figgap}a we show the results for the
gaps using $\Lambda=260$ meV, which at low $T$ are in good agreement with
the experimental results of Ref.\ \cite{ding}, even if a larger $T_c$ is
found here because fluctuation effects, that are expected to be important
in 2D, are not included. We point out that in a multiband system dominated
by off-diagonal (interband) coupling, deviations from the BCS value of
$2\Delta/T_c$ in each band can be obtained without invoking exotic
mechanisms. In Fig.\ \ref{figgap}a we find
$2\Delta_\beta/T_c=2$ and $2\Delta_\alpha/T_c=4$, and further enhancement
could be expected after the inclusion of fluctuation effects.

When multiple gaps are present, one may expect a signature on the
temperature behavior of the superfluid density $\rho_s$.  For a generic
lattice dispersion $\epsilon_\bk$, the superfluid density at BCS
level is given by:
\beqa
\label{rhosgen}
\rho_s&=&\int \frac{d^2\bk}{2\pi} \frac{\de^2\epsilon_\bk}{\de k_\alpha^2}
\left(1-\frac{\epsilon_\bk-\mu}{E_\bk} \tanh\left(\frac{
  E_\bk}{2T}\right)\right)+\nonumber\\
&+&2 \int \frac{d^2\bk}{2\pi} \left(
\frac{\de \epsilon_\bk}{\de k_\alpha}\right)^2
\frac{\de f(E_\bk)}{\de E_\bk},
\eeqa
where $f(x)=[1+\exp(x/T)]^{-1}$ is the Fermi function. Let us first
consider the electron bands $\gamma$: since they are almost empty, we can
approximate the full dispersion $\epsilon_\bk$ with the parabolic band
$\epsilon_\gamma$ introduced above, so that the previous equation reduces
to $\rho^\gamma_s={2t_\gamma n_\gamma} +{4t_\gamma N_\gamma} \int d\epsilon
\, \epsilon f'(E_\gamma)$.  As $T\rightarrow 0$ the quasiparticle
excitations vanish, due to the derivative $f'(E_\gamma)$, and the usual
result $\rho_s^\gamma=2t_\gamma n_\gamma $ is recovered.  For hole bands, a
similar expression can be used after a particle-hole transformation, which
insures that only the hole states at the
top of the band contribute to the superfluid density,
i.e. $\rho_s^{\alpha}= 2t_{\alpha}(2-n_{\alpha})$ at $T=0$.

%
\begin{figure}[htb]
\includegraphics[scale=0.3,angle=-90]{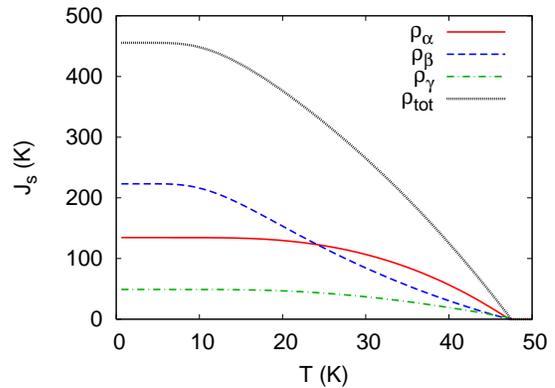}
\caption{(Color Online) Temperature dependence of the superfluid density in
  units of Eq.\ (\ref{js}) for the
  various bands.} 
\label{figsup}
\end{figure}
%

The results for the temperature dependence of the superfluid density for
the same parameters of Fig.\ \ref{figgap}a are shown in Fig.\ \ref{figsup},
where we plot the total $\rho_s$ together with the contributions from the
single bands.  In the $\beta$ band, which has the smallest gap, the clear
signature of the two different order parameters can be seen most clearly,
while the two other bands have a more standard behavior.  As we observed
above, for the $\beta$-band $2\Delta_\beta/T_c=2$, i.e. it is smaller than
the BCS value. This means that $\Delta(T)$ is finite and
sizable in a regime of temperature where it would be zero if not coupled
to another band.  As a consequence, at low $T$ $\rho_s^\beta$ decreases
quite rapidly, as it would vanish at a temperature proportional to
$\Delta_\beta$, while at higher temperature the depletion softens in a tail
which closes at the higher $T_c$ determined by the interband coupling. The
possibility to observe a clear change of curvature in the total superfluid
response $\rho_s^{tot}$ depends on the relative filling of the band with
the smaller gap. In Fig.\ \ref{figsup} we show that the presence of the
multiple gaps leads to deviations of $\rho_s^{tot}(T)$ with respect to the
single-band case, but with no upward curvature as in $\rho_s^\beta(T)$.  We
notice that $\rho_s^{tot}$ strongly resembles the one
reported recently in Ref.\ \cite{uemura_ba} in a
Ba$_{0.55}$K$_{0.45}$Fe$_2$As$_2$ compound with similar doping of the one
studied by ARPES in Ref.\ \cite{ding}.

More intriguing is instead the temperature dependence observed in several
electron-doped LaO$_{1-x}$F$_{x}$FeAs\cite{klauss_short,klauss_long} and
NdO$_{1-x}$F$_{x}$FeAs\cite{martin} compounds, where the measured $\rho_s$
resembles the one we find for the $\beta$ band. This effect is particularly
pronounced in LaO$_{1-x}$F$_{x}$FeAs for $x>0.1$, and suggests that the
smaller gap is formed on the electron part of the FS, which is expected to
give the larger contribution to the superfluid response in electron-doped
compounds.
Even though some indications of multiple gaps in
LaO$_{1-x}$F$_{x}$FeAs\cite{gonnelli} already exist, a clear experimental
observation of the gap opening on the different FS sheets is still lacking.

We finally comment on the $T=0$ value of the superfluid density, which is
proportional to the Fermi energy within our BCS scheme, and it is
therefore of the order of several hundreds of K. This energy scale is
comparable to the values currently available from penetration-depth
measurements in both hole-doped\cite{li_optics,uemura_ba} and
electron-doped\cite{klauss_short,klauss_long,martin} oxypnictides. To make
a direct comparison with the experiments, we recall that in two dimensions
the quantity $J_s=\hbar^2n_s^{2d}/m$ is an energy, where $n_s^{2d}$ is a
two-dimensional density of superfluid carriers. By considering each plane
as the basic 2D unit, one can relate $n_s^{2d}$ and $J_s$ to the measured
penetration depth $\lambda$ as:
\beq
\label{js}
J_s=\frac{\hbar^2 n_s^{2d}}{m}=\frac{\hbar^2 c^2}{4\pi
  e^2}\frac{d}{\lambda^2}=2.48 \frac{d[\AA]}{\lambda^2[(\mu m)^2]} K
\eeq
where $d$ is the interlayer spacing, which ranges between 4.3-6.6 \AA \ in
oxypnictides. The penetration depth of a Ba$_{0.6}$K$_{0.4}$Fe$_2$As$_2$
sample has been estimated in Ref.\ \cite{li_optics} from infrared
measurements as $\lambda=0.208$ $\mu$m. This gives according to Eq.\
(\ref{js}) $J_s\simeq 380$ K, which is not much lower that the value we
estimated above assuming the clean limit. Moreover, since the
optical-conductivity measurements of Ref.\ \cite{li_optics} 
suggest that the sample is in the dirty limit, we would expect that only a
fraction of the $\rho_s$ estimated from the (renormalized) band structure
will actually condense in the SC phase. We can then conclude that the
measured superfluid-density values are quite consistent with band-structure
calculations. For this reason, the observed similarity of the relation of
$1/\lambda^2$ vs $T_c$ in oxypnictides and cuprates
superconductors\cite{klauss_short} should {\em not} be taken as a signature
of similar unconventional behavior in the two classes of materials. Indeed,
in cuprate superconductors the number of carriers is large, because one is
doping the system with respect to half-filling, where the superfluid
density is expected to attain its maximum value in a hypothetical
uncorrelated system. In other words, one would
expect $\rho_s\sim (1-x)t$, while one measures values of order $\rho_s\sim
xt$, where $x$ is the doping and $t$ the typical hopping parameter. For
this reason, correlation effects as the proximity to a Mott
insulator\cite{massimo_rhos} or enhanced phase fluctuations\cite{lara_pp}
have been invoked as possible explanation of the low superfluid density
in cuprates. Instead, in oxypnictides the system has a low superfluid
density because an almost full hole band has a
vanishing contribution to $\rho_s$, and one expects $\rho_s\sim xt$
which is indeed near to the experimental values.

In conclusion, we have presented calculations in a four-band model built on
the basis of the experimental results on BaFe$_2$As$_2$, assuming that the
pairing arises from the interaction between hole-like and
electron-like Fermi surfaces, with different strengths due to
different nesting properties of the various FS sheets.  Within our
multiband BCS approach we determine parameters that account for the
multiple gaps recently observed by Ding {\sl et al.}\cite{ding}. Using the
same parameters, we compute the superfluid density, which does not show
striking signatures of the multiple gaps, in agreement with
experiments. The same agreement holds for its zero-temperature value, which
scales with the small density of carriers. This has to be contrasted
to the case of cuprates, where the smallness of $\rho_s$ is associated to
correlation effects.


We acknowledge R.~Arita, L.~Boeri, X.~Dai, and M.~Grilli for useful
discussions.

\end{document}